\documentclass{cimento}

\usepackage{graphicx}
\usepackage{soul}

\title{Optical design of the Tor vergata Synoptic Solar Telescope (TSST)}
\author{G.~Viavattene\from{ins:x}\thanks{currently at INAF - Osservatorio Astronomico di Roma, Via Frascati 33, 00078 Monte Porzio Catone, Italy},\ETC on behalf of 
D.~Calchetti\from{ins:x},
F.~Berrilli\from{ins:x},
D.~Del Moro\from{ins:x},
L.~Giovannelli\from{ins:x},
E.~Pietropaolo\from{ins:y},
M.~Oliviero\from{ins:z},
        \atque
L.~Terranegra\from{ins:z}
}
\instlist{\inst{ins:x} Universit\`a degli Studi di Roma ``Tor Vergata'', Via della Ricerca Scientifica, 1, 00133 Rome, Italy
\inst{ins:y} Universit\`a degli Studi dell'Aquila, Via Vetoio, 1, 67100 Coppito (AQ), Italy\\
 \inst{ins:z} INAF Osservatorio Astronomico di Capodimonte, Via Moiariello 16, 80131 Naples, Italy\\
  
  }

\begin{document}

\maketitle

\begin{abstract}
Synoptic full-disk solar telescope are fundamental instruments for present and future Solar Physics and Space Weather. They are typically used to study and monitor the solar activity by using high temporal cadence observations at different wavelength. The TSST (Tor vergata Synoptic Solar Telescope) is a new synoptic telescope composed of two spectral channels: an H$\alpha$ (656.3 nm) telescope and a Magneto Optical Filter (MOF)-based telescope in the Potassium (KI D1) absorption spectral line at 769.9 nm. H$\alpha$ observations are fundamental for the identification of flaring regions. The MOF-based telescope will produce line of sight magnetograms and dopplergrams of the solar photosphere, which are respectively used to study the magnetic field's geometry in active regions and dynamics of the solar atmosphere. In this work, we present an overview on the TSST and the optical design and characteristics of the MOF-based telescope, whose optical scheme is a double-Keplerian 80mm refractor with an aberration-free imaging lens.\\
\end{abstract}

\section{Introduction}

The Tor vergata Synoptic Solar Telescope (TSST) is part of a collaboration between the MOTH (Magneto Optical filters at Two Height) telescope \cite{Moretti2007,forte2018,Jefferies2019}, developed at IfA-University of Hawaii (Georgia State University and JPL), and the VAMOS (Velocity And Magnetic Observations of the Sun) telescope \cite{oliviero2010}, developed at the INAF Observatory of Capodimonte (Naples, Italy). This kind of synoptic telescopes are used for the simultaneous acquisition of line-of-sight (LoS) magnetograms and dopplergrams, provided by a tandem of Magneto-Optical Filters (MOFs), at multiple altitudes of the solar photosphere/chromosphere. More in detail, this prototypal network produces synoptic full-disk (spatial resolution of 4 arcsecs), or intermediate resolution (1 arcsecs), solar LoS magnetograms and dopplergrams with a temporal resolution of 5 seconds, faster than existing solar facilities (HMI \cite{scherrer2012}, GONG \cite{harvey1996}) which have a temporal cadence from 30 seconds to 2 minutes, relevant to Space Weather and astrophysical studies.\\
In this work, after introducing the theory behind the MOF operation, we describe the TSST and we present its polarimetric and the optical scheme. The optical scheme has been implemented using Zemax\copyright\ as optical design software.\\

\section{The Magneto-Optical Filters theory}

\begin{figure}[t]
\centering
\includegraphics[width=0.6\textwidth]{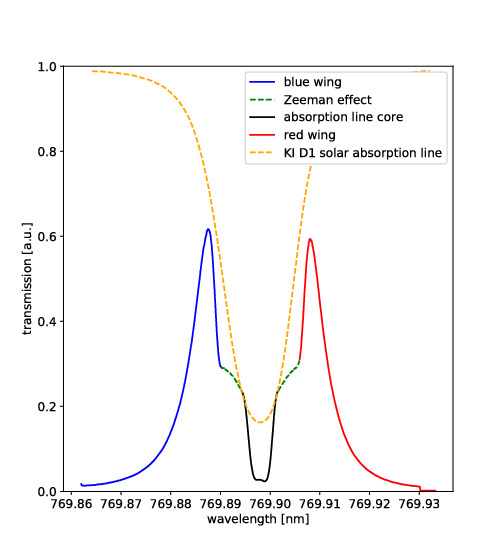}
\caption{Spectral transmission of the MOF cell around the Potassium KI D1 spectral line at 769.9 nm. The blue and red wings are plotted respectively in blue and red. The Zeeman effect near the line core is plotted with green dashed lines. The line core is plotted in black. The reference Potassium KI D1 spectral line from atlas is plotted with yellow dashed line. Figure from \cite{calchetti2020}.}
\label{FIG1}
\end{figure}

A typical MOF \cite{cimino1968,agnelli1975} is a cell containing a vapor of an alkalin metal placed inside a strong longitudinal magnetic field and between two crossed linear polarizers (P1 and P2). Available MOF cells are filled with sodium, potassium, calcium \cite{rodgers2005} or helium vapor \cite{murphy2005} and are suitable for space applications \cite{Berrilli2010,Moretti2010}.\\
The MOF cells are made by cylindrical glass tubes and their extremities are closed by two plano-parallel and flat optical window ($\lambda$/4 optical quality). On the lateral surfaces of the cylindrical cells, two wells are located and they are filled with an alkaline solid metal (in our case Potassium). The alkaline vapor is generated by heating these two wells using a Ni-Cr wire, with a total resistance of 20 $\Omega$, powered by 24 V DC. The Potassium condensation on the two optical windows is avoided by filling the cells with Argon gas.\\
The longitudinal magnetic field B is generated by a powerful (about 0.1-0.2 tesla) magnetic circuit, using rare earth magnets, which causes a Zeeman splitting of the Potassium KI D1 spectral lines at 769.9 nm. Therefore, only the two $\sigma_{+}$ and $\sigma_{-}$ components should emerge from the cell with left and right circular polarization respectively; the linear polarized $\pi$ component is not visible because it oscillates along the optical axis. The spectral separation between the two components can be calibrated by varying the magnetic field intensity. Inside the cell, the Macaluso-Corbino quantum effect happens \cite{macaluso1898,macaluso1899a,macaluso1899b}, which causes the rotations of the linear polarization on the wings of the two $\sigma_{+}$ and $\sigma_{-}$ components. The Macaluso-Corbino effect is similar to the Faraday rotation, but it happens only in the proximity of resonance spectral lines.\\
Without these effects we would not see light passing through the pair of crossed linear polarizers
P1 and P2 (except for the radiation leakage due to the non-ideal behavior of the crossed linear polarizers or due to possible errors in the crossing of the polarizers). Instead, circular polarized light, in correspondence of $\sigma_{+}$ and $\sigma_{-}$, and rotated linearly polarized light arrive on P2 thanks to the Zeeman and Macaluso-Corbino effects on the vapor atoms. Therefore, the transmission profile of a MOF cell will contain two peaks of transmission on the external wings of the Zeeman splitted lines. In Fig. \ref{FIG1}, we report the spectral transmission profile of our spectral MOF cell in the Potassium KI D1 spectral line at 769.9 nm \cite{calchetti2020}. Varying the heating temperature of the two wells, and so the vapor density inside the cell, the amplitude and the wavelength of the two Macaluso-Corbino peaks can be varied.\\
Finally, by measuring both polarization states (+ and -) in the red (R) and blue (B) wings of the observed solar absorption spectral line LoS dopplergrams (D) and magnetograms (M) of the whole solar disk can be computed \cite{forte2018} using the following relations:

\begin{equation}
D=\frac{R^+-B^+}{R^++B^+}+\frac{R^--B^-}{R^-+B^-}
\label{EQ1}
\end{equation}{}
\begin{equation}
M=\frac{R^+-B^+}{R^++B^+}-\frac{R^--B^-}{R^-+B^-}
\label{EQ2}
\end{equation}{}

The R and B peaks of the Macaluso-Corbino effect must be spectrally selected to measure the signals in the two wings of the KI D1 spectral line. A second cell, filled with Potassium vapor, is necessary for this purpose. The first cell is the MOF spectral cell and the second cell is called Wing Selector (WS). Therefore, the WS cell absorbs the radiation at only one of the Zeeman component, letting the radiation pass at the other Zeeman component. The magnetic field of the WS cell must be higher than the one of the MOF cell because its Zeeman component must be superimposed to the transmission peaks of the Macaluso-Corbino effect of the MOF cell. Tuning the magnetic field and the heating temperature of the wells of the WS cell, and rotating alternatively the axis of an half wave plate (between 0$^{\circ}$ and 45$^{\circ}$) used in combination with a quarter wave plate (set at 45$^{\circ}$), it is possible to select alternatively one of the two transparency peaks of the Macaluso-Corbino effect.\\

\section{Tor vergata Synoptic Solar Telescope}

The TSST is a double-channel synoptic solar telescope made by two optical telescopes for Space Weather applications which provides H$\alpha$ images, dopplergrams and magnetograms of the solar atmosphere. It will be one of the telescopes involved in a network of solar observatories, including also the MOTH and the VAMOS telescopes, with the aim to obtain multi-height observations of the Sun.\\
The two telescopes of the TSST will be mounted on an EQ8 PRO Synscan equatorial mount by Sky-Watcher, which has a load capacity of 50 kg. The H$\alpha$ telescope is a Daystar SR-127, an achromatic refractor with 127 mm aperture and 4064 mm of focal length. This long focal length is reached using an internal 4X telecentric Barlow lens. In the collimated beam between the two components of the Barlow lens, a temperature-stabilized Fabry-Perot Interferometer is placed (classical mount) and it is used to select the H$\alpha$ spectral line of the solar spectrum. The bandpass of the Daystar SR-127 is 0.04 nm, making sure to see with high contrast the flaring active regions and filaments. The H$\alpha$ telescope will be equipped with a 0.33X focal reducer (1341 mm of effective focal length) in order to have an image of the whole solar disk of 12.7 mm on the focal plane. The two crown-facing identical achromatic doublets of the focal reducer guarantee its aberration control (the aberrations introduced by one doublet are removed by the other) and produce a reducing factor of 0.5X; the reducing factor becomes 0.33X inserting a spacing of 20mm between the focal reducer and the focal plane camera.\\
The MOF telescope has been entirely designed with Zemax\copyright\, software and it has an 80 mm aperture and an effective focal length of 963 mm, in order to have a focal plane image comparable to the one retrieved by the H$\alpha$ telescope. Its optical scheme and operation will be treated in detail in the next Section. This telescope will provide magnetograms and dopplergrams of a solar atmospheric layer at approximately 300$\div$400 km above the base of the solar photosphere, which is the layer where the Potassium KI D1 spectral line at 769.9 nm is mainly formed.\\
The two telescopes will be equipped with two identical CMOS Dalsa Pantera 1M60 cameras during the preliminary phase. The camera sensor has 1024$\times$1024 pixels of 12 $\mu$m of dimension (12.3$\times$12.3 mm sensor dimension). The pixel scale of the H$\alpha$ and the MOF telescopes are respectively 1.8 and 2.6 arcsec/pixel. The camera quantum efficiency for the H$\alpha$ spectral line at 656.3 nm and for the Potassium KI D1 spectral line at 769.9 nm is approximately 50\%.\\

\section{Optical design of the MOF-based telescope}

\subsection{Polarimetric scheme}

The polarimetric scheme of the MOF channel of the TSST is similar to the one used in the VAMOS telescope \cite{oliviero1998a,oliviero1998b,oliviero1998c,vogt1999,oliviero2002}. Referring to Fig. 1 in \cite{calchetti2020}, we describe the polarimetric scheme of the MOF channel. After the first optics and filters, the first polarimetric element reached by the solar light is the first quarter wave plate (QWP), with its axis set at 45$^\circ$ with respect to the axis of the following first half wave plate (HWP), which has, in its turn, the axis parallel to the axis of the P1 linear polarizer. The alternative selection of the two left and right circularly polarized components of the incoming solar light is performed by rotating the axis of the first HWP between 0$^\circ$ and 45$^\circ$ with respect to the axis of P1. Therefore the two components pass alternatively through P1 and go inside the MOF cell, since their polarization axes have been directed parallelly or perpendicularly to the axis of P1 by the first HWP. The axis of the linear polarizer P2 is perpendicular to the axis of P1. The axis of the second HWP is set parallel to the axis of P2. The axis of the second QWP is at 45$^\circ$ with respect to the axis of the second HWP. Rotating the axis of the second HWP between 0$^\circ$ and 45$^\circ$, the radiation coming from the MOF cell possesses alternatively left and right circular polarization state. Then, the WS cell selects alternatively one of the two transmission peaks of the MOF cell. At the end of the polarimetric section, the solar light is focused on the final image plane.\\

\subsection{Optical scheme}

We design the entire optical scheme of the MOF channel using Zemax\copyright\, optical design software. The optical scheme has been implemented in order to satisfy the following requirements and constraints:

\begin{itemize}
\item Refractor telescope, in order to avoid the pupil obstruction produced by the secondary mirror of a Newtonian-based or Cassegrain-based reflective telescope;
\item Entrance pupil diameter of 80 mm, in order to have an angular resolution of 2 arcsec (corresponding to 1.5 Mm of spatial resolution) on the solar photosphere;
\item Large Field-of-View (FoV) in order to perform full disk imaging, at least 0.25 degrees of half field cone angle;
\item Collimated beam completely inside the MOF and the WS cells (without vignetting), with the pupils inside the cells, in order to avoid the internal vapor seeing;
\item Collimated beam in the polarimetric optics (HWPs, QWPs and linear polarizers), in order to have a better spectral response;
\item High-pass and low-pass filters to remove the unwanted solar radiation from the optical path in order to avoid the overheating of the optics;
\item Aberration free image of the whole solar disk using corrective lens near the final image focal plane;
\item Compact scheme to reduce the encumbrances and weights for the EQ8 SynScan equatorial mount, considering also the weight of the H$\alpha$ telescope.
\end{itemize}

\begin{figure}[h]
\centering
\includegraphics[width=1.25\textwidth, angle=270]{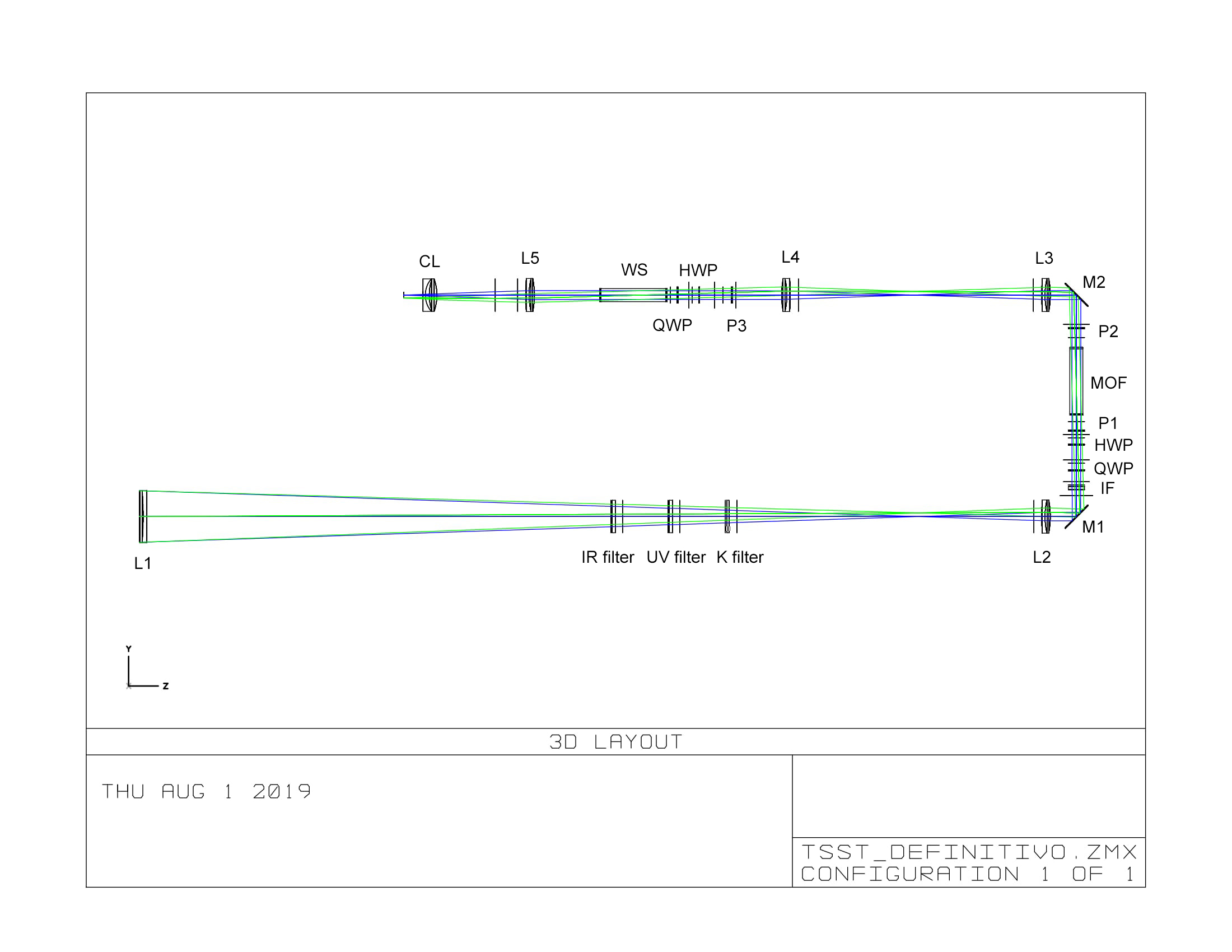}
\caption{Optical scheme of the MOF TSST telescope. L1 is the objective lens. L2, L3, L4 and L5 are identical achromatic doublet lens. M1 and M2 are flat folding mirrors. QWP and HWP are respectively quarter wave plates and half wave plates. P1, P2 and P3 are linear polarizer. CL is a group of corrective lens.}
\label{FIG2}
\end{figure}

\begin{figure}[h]
\centering
\includegraphics[width=0.8\textwidth]{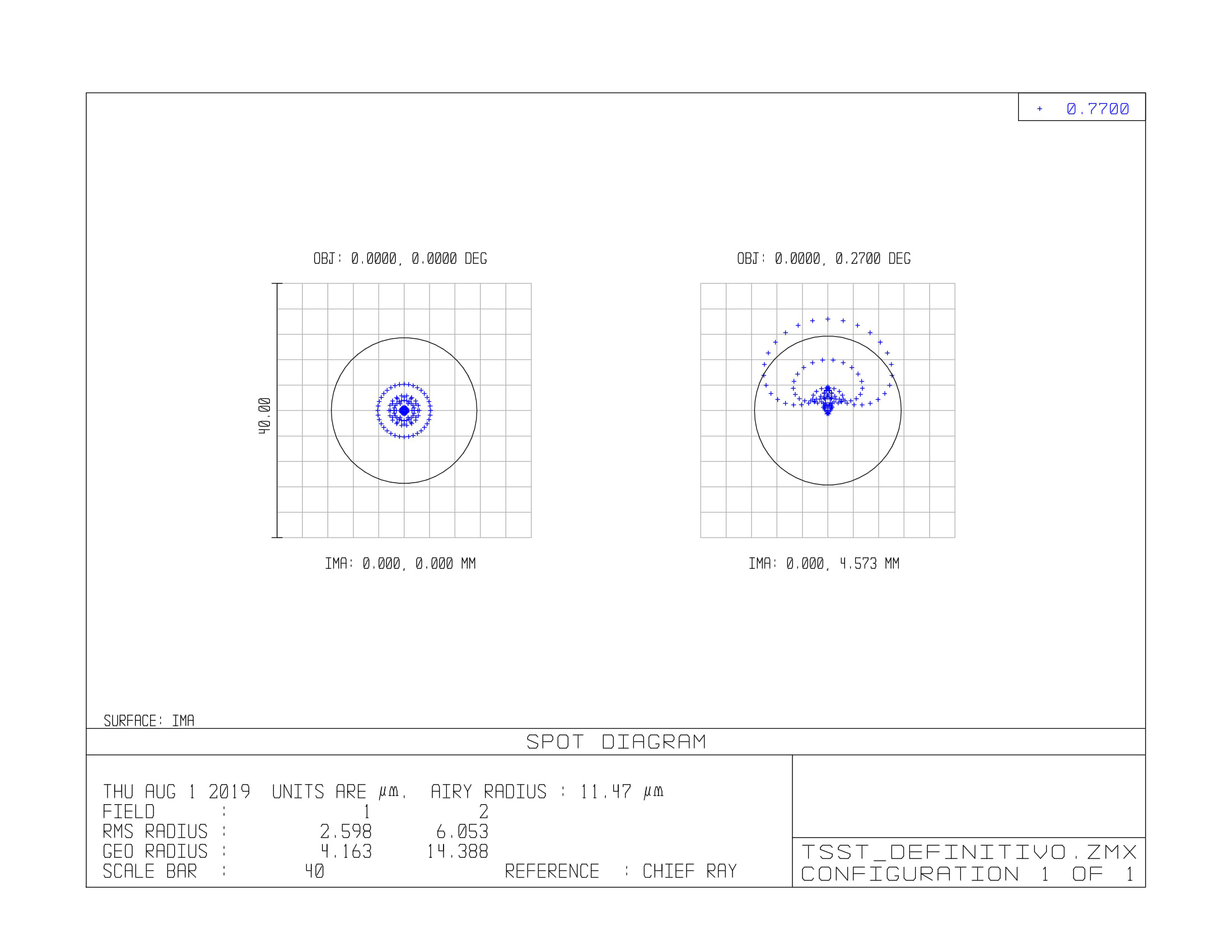}
\caption{Spot diagrams in the image focal plane of the MOF channel of the TSST at the center (on the left) and at the edge (0.27 degrees, on the right) of the FoV. The airy disk is shown with a black circle.}
\label{FIG3}
\end{figure}

Several optical schemes have been proposed and studied for the MOF telescope. The final scheme includes a couple of folding mirrors to reduce the total length of the instrument and keep total axial momentum within the limit of the equatorial mount. Folding mirrors 
can introduce errors to the polarization detecting system however these are negligible for our specific application. Furthermore, small inclination angles ($<$ 45$^{\circ}$) could introduce astigmatism. The adopted solution with two folding mirrors at 45$^{\circ}$ and and additional polarimeter meets these requirements.\\
The entire optical scheme of the MOF telescope is reported in Fig. \ref{FIG2}. The objective lens L1 is a \textit{Spindler \& Hoyer} achromatic doublet with a diameter of 80 mm and a focal length of 1185 mm. Three pre-filters are placed in the converging beam after L1: an high-pass IR cut filter ($>$ 825 nm), a low-pass UV cut filter ($<$ 700 nm) and a broadband Potassium filter (41 nm of bandwidth centered at 769 nm). These three pre-filters are used to dominate the overheating of the optics, to reject the unwanted and useless solar radiation and to avoid the Intensity Effect in the MOF cell \cite{oliviero2011} (saturation effect of the Potassium vapor atoms due to excessive radiation which modifies the spectral behaviour of the MOF cell due to decrease of the vapor optical depth). L2, L3, L4 and L5 are identical achromatic doublets (diameter 50.8 mm, focal length 200 mm and anti-reflection coating from 650 nm to 1050 nm); they are oriented in order to realize a double-Keplerian telescope with an imaging lens. M1 and M2 are two identical flat folding mirror (diameter 50.8 mm, optimized for the near IR) used to fold the optical beam and to reduce the dimension of the MOF telescope. L2 is placed after the image produced by L1 and it is used to collimate the solar beam for the first polarimetric block and for the MOF cell. After M1, which folds the optical path, an interference filter (IF, 2 nm of bandwidth) centered in the Potassium spectral line at 769.9 nm is placed with the aim of reducing the spectral straylight and the spurious images produced by the various components of the optical system and to increase the contrast of the image. The first QWP, the first HWP, the linear polarizer P1, the MOF cell and the linear polarizer P2 are still placed in the collimated beam. M2 is used to fold again the optical beam. The lens L3 creates a second image plane and the lens L4 collimates again the solar beam for the second polarimetric block and for the WS cell. An additional linear polarizer P3 is used to cancel the spurious polarization signal that could be introduced by M2. Still in the collimated part, the second HWP, the second QWP and the WS cell are placed. L5 creates the final image focal plane. Between L5 and the final focal plane, a corrective lens (CL) system is placed to correct the focal plane aberrations, especially the field curvature and distortion, induced by the high focal ratio (f/4) of the achromatic doublets. The CL system is made by a bi-convex N-BK7 lens (diameter 50.8 mm, focal length 75 mm) and a plano-concave N-BK7 lens (diameter 50.8 mm, focal length -75 mm). The mutual distance between the two corrective lenses, their distances from L5 and their distance from the final focal plane have been optimized with Zemax\copyright\, in order to have the best aberration-free image.\\
The spot diagrams in the final focal plane of the MOF telescope at the center and at the edge of the FoV (0.27 degrees of half field cone angle) are shown in Fig. \ref{FIG3}, with the Airy diffraction disk plotted as a black circle. As it can be seen from the figure, the image is diffraction limited at the center of the FoV and almost diffraction limited at the edge of the FoV. The image at the edge of the FoV will not be appreciably degraded by the rays that go outside the Airy disk because they will be smeared by the typical diurnal seeing (1-2 arcsec) and because they are smaller than the image pixel scale reached by the Dalsa Pantera 1M60 CMOS camera.\\
The ray fan plots, which report the rays positions in the image plane with respect to the ones in the pupil plane, in the final focal plane of the MOF telescope at the center and at the edge of the FoV (0.27 degrees of half field cone angle) are shown in Fig. \ref{FIG3}. It can be seen that there is a residual spherical aberration and even less distortion at the center of the FoV, and a residual distortion and even less spherical aberration and field curvature at the edge of the FoV. Despite these residual aberrations, we can declare that with good approximation the MOF telescope of the TSST is almost diffraction limited and that it works in optimal conditions for solar diurnal observations.\\
The tolerance analysis performed with Zemax\copyright\, allowed us to estimate that maximum misalignment of 0.2 mm and maximum tilts of 1 degrees will not affect the telescope performances, preserving its diffraction limit behavior.\\

\begin{figure}[h]
\centering
\includegraphics[width=0.8\textwidth]{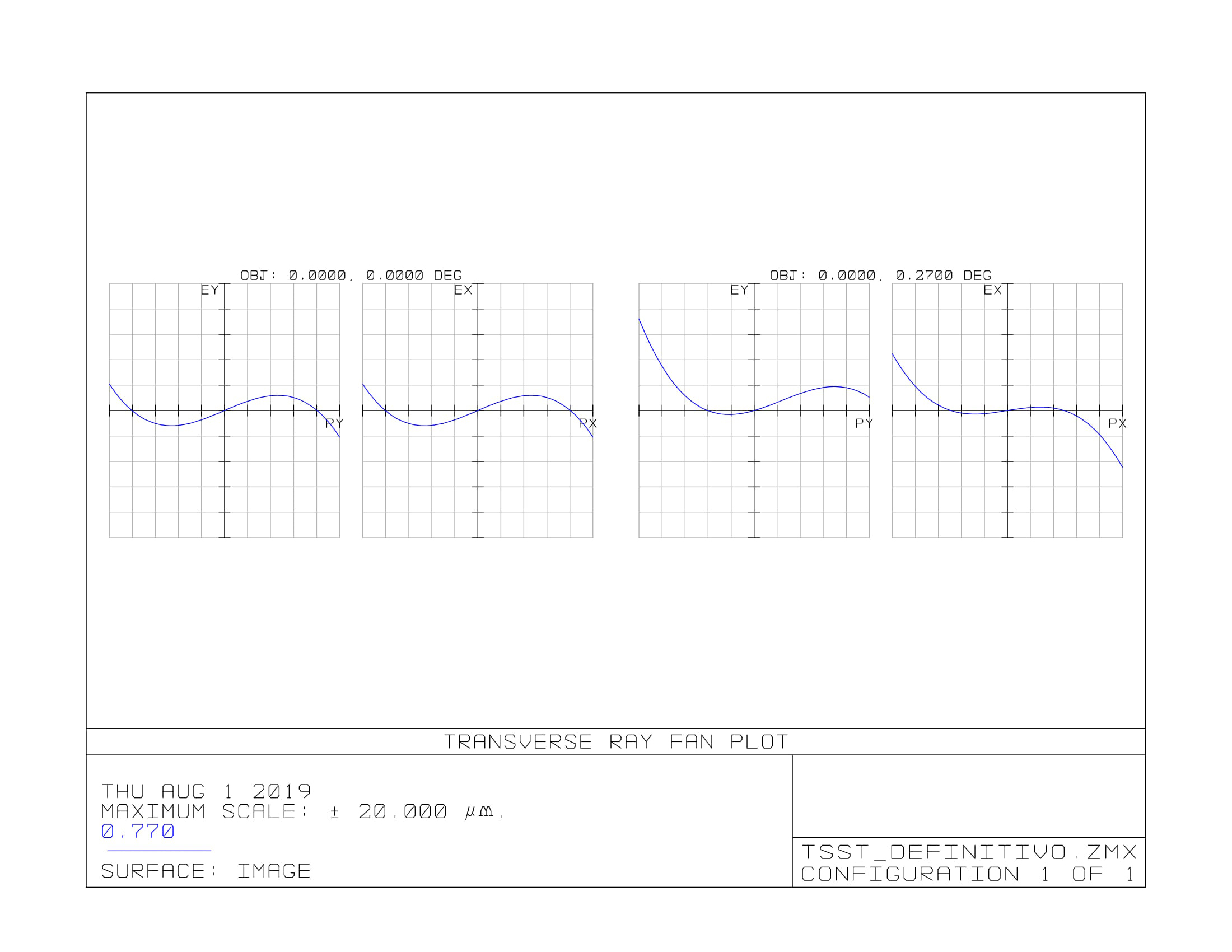}
\caption{Ray fan plot in the image focal plane of the MOF channel of the TSST at the center (on the left) and at the edge (0.27 degrees, on the right) of the FoV.}
\label{FIG4}
\end{figure}

\section{Conclusions and future perspectives}
\begin{figure}[h]
\centering
\includegraphics[width=0.95\textwidth]{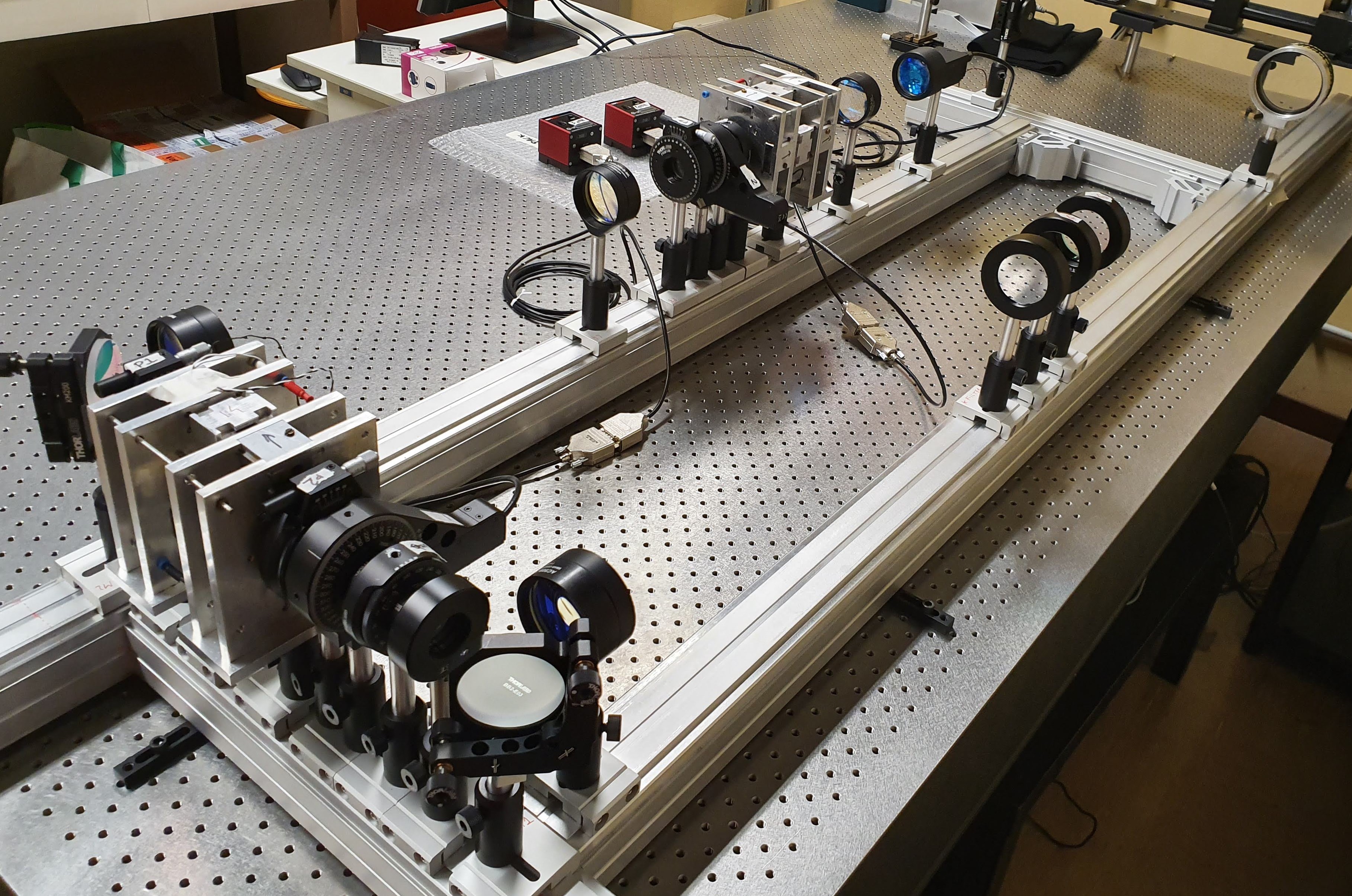}
\caption{MOF-based telescope of the Tor vergata Synoptic Solar Telescope in assembly phase. All the optical components have been mounted on the rails, aligned and collimated.}
\label{FIG5}
\end{figure}

Solar synoptic telescopes are fundamental for Space Weather applications. Full-disk H$\alpha$ images, magnetograms and dopplergrams are respectively used for the detection and identification of flaring active regions, for characterize the magnetic fields geometry and to study the dynamics of the solar atmosphere \cite{berrilli2018,falconer2011}. Here, we presented the  Tor vergata Synoptic Solar Telescope, a double-channel H$\alpha$ and MOF-based solar telescope. We describe in detail the optical scheme and the operation of the MOF-based telescope, which is devoted to the full-disk magnetograms and dopplergrams acquisition, and we discussed its optical performances. Currently, the MOF channel of the TSST is in assembly phase and its optical elements have been recently aligned and collimated (see Fig. \ref{FIG5}). The spectral characterization of the instrument has benn obtained at the MOF spectral facility of the INAF-OAC in Naples using an high-precision tunable laser and the first light is planned in the second half of 2020. The robotic and remote control of the whole system will be realized in collaboration with Department of Physical and Chemical Science (DSFC) in L'Aquila.

\acknowledgments
The author would like to thank the Organizers of the 105th National Congress of the Societ\`a Italiana di Fisica (SIF), held in L'Aquila (Italy) in September 2019, for the Special Mention of the Astrophysics Session for the talk related to this work. This research is partially supported by the Italian MIUR-PRIN grant 2017APKP7T on Circumterrestrial Environment: Impact of Sun-Earth Interaction.\\

\end{document}